\documentclass[aps,prd,dlspace,twocolumn]{revtex4-1}
\usepackage{epsfig}
\usepackage{appendix}
\usepackage{graphics}
\usepackage{graphicx}
\usepackage{color}
\newcommand{\jpsi}{J / \psi}

\newcommand{\old}[1]{}

\newcommand{\be}{\begin{equation}}
\newcommand{\ee}{\end{equation}}
\newcommand{\ba}{\begin{eqnarray}}
\newcommand{\ea}{\end{eqnarray}}
\newcommand{\bi}{\begin{itemize}}
\newcommand{\ei}{\end{itemize}}

\begin{document}
\title{Dissociation of heavy quarkonium in hot QCD medium in a quasi-particle model}
 \author{Vineet Kumar Agotiya $^{a}$}
\email{agotiya81@gmail.com}
\author{Vinod Chandra$^{b}$}
\email{vchandra@iitgn.ac.in}
 \author{M. Yousuf Jamal$^{b}$}
\email{mohammad.yousuf@iitgn.ac.in}
 \author{Indrani Nilima$^{a}$}
\email{nilima.ism@gmail.com}
 \affiliation{$^a$Centre for Applied Physics, Central University of Jharkhand
 Ranchi, India, 835 205}
 \affiliation{$^b$ Indian Institute of Technology Gandhinagar, Gandhinagar-382355, Gujarat, India}
\begin{abstract}
Following a recent work on the effective description of the  equations of state for hot QCD 
obtained from a Hard thermal loop expression for the gluon self-energy, in terms of the quasi-gluons and quasi-quark/anti-quarks 
with respective effective fugacities,  the  dissociation process of heavy quarkonium in hot QCD medium has been 
investigated.  This has been done by 
investigating the medium modification to a heavy quark 
potential. The medium modified potential has a quite different form 
(a long range Coulomb tail in addition to the usual Yukawa term) in contrast 
to the usual picture of Debye screening. The flavor dependence of the
binding energies of the heavy quarkonia states and the dissociation temperature  have been obtained
by employing the debye mass for pure gluonic and full QCD case computed employing the quasi-particle picture. 
Thus  estimated  dissociation patterns of the charmonium and bottomonium states, considering Debye mass from different approaches
 in pure gluonic case and full QCD,  have shown  good agreement with the other potential
 model studies.
\end{abstract}
\maketitle
\section{Introduction}
The problem of dissociation of bound states in a hot QCD medium is of great
importance in heavy ion collisions as it provides evidence for the creation of
the quark-gluon plasma there ~\cite{Leitch}. Matsui and Satz~\cite{matsui}, proposed  $J/\psi$ 
suppression caused by the Debye screening by the quark-gluon plasma (QGP) as an important 
signature to reaffirm its formation  in heavy ion collisions.
 The physical understanding of the  quarkonium dissociation within a deconfined medium
  has undergone some definite refinements in the last couple of years 
~\cite{Laine:2008cf,BKP:2000,BKP:2001,BKP:2002,BKP:2004}.
As the heavy quark and anti-quark in a quarkonia state  are bound together by almost static (off-shell) gluons, 
therefore,  the issue of their dissociation boils down to how the gluon self-energy behaves  at high temperatures.
It has been noticed that the  gluon self-energy has both real and imaginary parts~\cite{laine}. Note that 
the real part lead to the  Debye screening, while the imaginary part leads to 
Landau damping and give rise the thermal width to the quarkonia.

The fate of quarkonia, at zero temperature  can
 be understood in terms of non-relativistic potential models (as the velocity
 of the quarks in the bound state is small, $v\ll 1$) \cite{Lucha} using the Cornell potential \cite{Eichten}.
 Further, the physics of the fate of a  given quarkonium state in the QGP medium,  is encoded in its spectral
 function~ \cite{Iid06,Jak07}. Therefore, following the  temperature behavior of the spectral function,   theoretical insight to the
quarkonium properties at finite temperature can be made. There are mainly two lines of theoretical
approaches to determine quarkonium spectral functions, {\it viz.}, the potential models~\cite{potential2,Shu04,Won05,Cab06} 
which have been widely used to study quarkonia states
 (their applicability at finite temperature is still under scrutiny),
 and  the lattice QCD studies~\cite{lat_disso,haque} which provides
 the reliable way to determine spectral functions, but the results
 suffer from discretization effects and statistical errors, and thus are
 still inconclusive. These two approaches show poor matching as far as
 their predictions are concerned. None of these two approaches leads
 towards a complete framework to study the properties of quarkonia
 states at finite temperature. However, some degree of qualitative
 agreement had still  been achieved for the S-wave correlaters.
 In contrast, the finding was somehow ambiguous for the P-wave
 correlaters. Additionally, the temperature dependence of the potential
 model was even qualitatively different from the lattice one. 
Refinement in the computations of the spectral functions have
 recently been done (including the zero modes both in the S- and P-channels)~ \cite{Ume07,Alb08}. It has been observed that, these contributions cure
 most of the previously observed  discrepancies with lattice calculations.
 This supports the fact that the employment of potential models at finite
 temperature can serve as an important tool to complement lattice studies.
 The potential model can actually be derived directly from QCD as an effective
 field theory (potential non-relativistic QCD - pNRQCD) by integrating
 out modes above the scales $m_Q$ and then $m_Q v$, respectively
 \cite{Brambilla,kaka,pk}.  
 
 Note that the potential models have 
 been served as  useful approach while exploring the 
 physics of heavy quarkonia since the discovery of
 $J/\psi$ \cite{Brambilla,Escobedo:PRD2013}. It indeed provides a useful way
 to examine quarkonium binding energies, quarkonium wave functions,
 reaction rates, transition rates and decay widths. It further allows
 the extrapolation to the region of high temperatures by expressing
 screening effects reflecting on the  temperature dependence of the potential.
 The effects of dynamics of quarks on the stability
 of quarkonia can be studied by using potential models extracted from
 thermodynamic quantities that are computed  in full QCD. At high temperatures,
 the deconfined phase of QCD exhibits screening of static color-electric
 fields~\cite{GPY}; it is, therefore, expected that the screening will lead to
 the dissociation of quarkonium states.  After the success at zero
 temperature while predicting  hadronic mass spectra, potential model
 descriptions have also been applied to understand quarkonium properties
 at finite temperature. 
  
 Note that, the production of  $J/\psi$ and $\Upsilon$ mesons in hadronic
 reactions occurs in part via the production of higher
 excited $c \bar c$ (or $b \bar b$) states and their decay into
 respective ground state. Since the lifetime of different quarkonium
 state is much larger than the typical life-time of the medium  produced
 in nucleus-nucleus collisions; their decay occur almost completely
 outside the produced medium~\cite{lain,he1}. This is crucial due to
 the fact that the produced medium can be probed not only by the ground
 state quarkonium but also by different excited quarkonium states.
 Since, different quarkonia states have different sizes and
 binding energies, hence, one expects that higher excited states will
 dissolve at smaller temperature as compared to the smaller and more
 tightly bound ground state. These facts may lead to a sequential
 suppression pattern in $J/\psi$ and $\Upsilon$ yield in nucleus-nucleus
 collision as the function of the energy density. The potential model in
 this context could be helpful in predicting for the  binding energies
 of various quarkonia states by setting up and solving appropriate
 Schr\"{o}dinger  equation in the hot QCD medium. The first step
 towards this is to model an appropriate medium  dependent interquark
 interaction potential at finite temperature. The dissociation of
 heavy quarkonium derived by the presence of screening of static color
 fields in hot QCD medium has long been proposed as a signature of a deconfined
 medium, and QGP formation~\cite{matsui}. Since then, this has been
 an area of active research~\cite{nora, mocsy,umeda,patra,patra1,sdatta,nora1}. 
 However, a precise definition of the dissociation temperature is
 still elusive and is a matter of intense theoretical and
 phenomenological investigations either from the
 perspective of lattice spectral function studies~\cite{sdatta,spectral,petre,spt1,spt2} or potential inspired models~\cite{alberico,pot1,pot2,mocsy1} or
 effective quarkonia field theories~\cite{eff}. The heavy quarks/antiquarks,
 such as $c\bar{c}$ are bound together by almost static
 gluons~\cite{laine,dum,laine1}. Therefore, the gluon self-energy in the 
 static limit can be helpful in understanding the fate of such states
 in the hot QCD medium. 

While modeling the medium modified potential the non-perturbative
 effects coming from the non-zero string tension between the quark-antiquark
 pair in the QGP phase is not an unreasonable consideration. This is
 simply due to the fact that the hadronic to the QGP transition
 is a crossover. Therefore the string-tension  will not vanishes abruptly 
at or closer to $T_c$. One should certainly study its effect
 on the behavior of quarkonia even above the deconfinement temperature.
 This fact has been exploited  in the  recent past in Refs.~\cite{akhilesh,patra},  
 where a  medium-modified form of the heavy quark potential has been
 obtained by correcting the full Cornell potential, not only its
 Coulomb part alone,  as usually done in the literature, with a
 dielectric function encoding the effects of the deconfined medium.
The medium modified potential, thus obtained has a long-range Coulomb
 tail with an (reduced) effective charge~\cite{patra} along with 
the usual Debye-screened form employed in most of the literature.
 We subsequently used this form to  determine the binding energies and the 
dissociation temperatures of the ground and the first excited states of 
charmonium and bottomonium spectra. 

In the present paper, we shall consider an isotropic QGP
 medium which is described in terms of quasi-particle degrees of
 freedom based on a recently proposed quasi-particle model for hot QCD
 equations of state based on improved perturbative techniques at weak coupling~\cite{chandra1,chandra2}. We further implement the 
 similar description for the lattice QCD based equations of state~\cite{chandra_quasi}.  We first obtain the medium
 modified  heavy quark potential (both real and imaginary parts) and estimate
 the dissociation temperatures for 2-, and 3-flavor hot QCD medium.   As an intermediate step, the binding energies  of the different
 quarkonia state and  their respective thermal width have been obtained in the  Hot QCD/QGP medium. Our predictions have been found to be consistent to
 the results obtained from other approaches.

 The manuscript is organized as follows. The real part of the heavy-quark
 potential is discussed in Section II along with Debye mass obtained
 from a quasi-particle model of hot QCD equation of state along
 with binding energies of various quarkonia bound state by solving
 the Schr\"{o}dinger equation (numerically). In Section III, computations 
 on the imaginary part of the potential and thereby thermal width of the
 quarkonium has been presented. Section IV, deals with results and discussions. Finally, the conclusions and future prospects of the work has been
 presented in Section V.

\section{Heavy-quark potential}
The interaction potential between a heavy quark and antiquark gets
modified in the presence of a medium. The static interquark potential  plays 
vital role in understanding the fate of quark-antiquark bound states in the
 hot QCD/QGP medium. These aspects have been well studied in the 
 literature  and in this direction  several excellent reviews
 exist~\cite{Bram,kluberg} that covers potential model based 
phenomenology as well as on the lattice QCD based approaches. 
In all these studies, the form of the potential in the deconfined phase is 
of Yukawa form (screening coulomb). The prime assumption is that the
 melting of the string between the quark-antiquark pairs in the deconfined
 phase is motivated by the fact that there is a phase transition
 from a hadronic matter to a QGP phase. In the present analysis, we
 incorporate the modification to both the Coulomb part and 
confining part in the deconfined medium~\cite{patra,Petreczky:2005bd}.
 This is based on the fact that the transition between the hadronic to
 the QGP phase is  a cross-over as shown by the recent
 lattice studies~\cite{Rothkopf}. In the case of finite-temperature QCD we here employ the 
 Ansatz that the medium modification enters in the  Fourier transform of heavy quark
 potential, $V(k)$  as~\cite{patra} 
\begin{equation}
\label{eq3}
\tilde{V}(k)=\frac{V(k)}{\epsilon(k)} \quad ,
\end{equation}
where $\epsilon(k)$ is the dielectric permittivity which is obtained from the static limit of the longitudinal part of gluon 
self-energy\cite{schneider}
\begin{eqnarray}
\label{eqn4}
\epsilon(k)=\left(1+\frac{ \Pi_L (0,k,T)}{k^2}\right)\equiv
\left( 1+ \frac{m_D^2}{k^2} \right).
\end{eqnarray}

In our case,  $V(k)$ in Eq.(\ref{eq3}) is the Fourier transform (FT) of
the Cornell potential (to compute the FT we need to introduce a modulator of the form $\exp(-\gamma r)$ and finally let the $\gamma$ tends to zero),  which is obtained as~ 
\begin{equation}
\label{eqn5}
{\bf V}(k)=-\sqrt{(2/\pi)} \frac{\alpha}{k^2}-\frac{4\sigma}{\sqrt{2 \pi} k^4}.
\end{equation}
Next, substituting Eq.(\ref{eqn4}) and Eq. (\ref{eqn5}) into Eq. (\ref{eq3}) and 
 evaluating the inverse FT, we obtain r-dependence of the
 medium modified potential~\cite{akhilesh,ldevi}:
\begin{eqnarray}
\label{eq6}
{\bf V}(r,T)&=&\left( \frac{2\sigma}{m^2_D}-\alpha
\right)\frac{\exp{(-m_Dr)}}{r}\nonumber\\
&-&\frac{2\sigma}{m^2_Dr}+\frac{2\sigma}{m_D}-\alpha m_D
\end{eqnarray}
Interestingly, this potential has a long range Coulombic tail in
 addition to the standard Yukawa term. The constant terms are introduced
 to yield  correct limit of $V(r,T)$ as $T\rightarrow 0$
 (it reduces to the Cornell form). Note that such terms could appear
 naturally while performing the basic computations of real time static
 potential in hot QCD~\cite{const1} and from the real and imaginary
 time correlators in a thermal QCD medium~\cite{const2}. The three
 dimensional form is motivated from the fact that at finite temperature,
 the flux tube structure may expand in more than one dimension~\cite{hsatz}.
In  the limiting case  $r>>1/m_D$, the dominant terms in the potential are
 the long range Coulombic tail and  $\alpha m_D$. The potential will look as,
\begin{eqnarray}
\label{lrp}
{V(r,T)}\sim -\frac{2\sigma}{m^2_Dr}-\alpha m_D
\end{eqnarray}, 
and can be tackled analytically while solving for the  binding energies
 and the dissociation temperatures for the ground and first excited
 states of $c \bar c$ and $ b \bar b$. In general, one require to set
 the Schr\"{o}dinger equation with the full potential and solve it
 numerically for the binding energy. Here, we consider the full potential 
 and estimate the binding energies and the dissociation temperatures for heavy quarkonia. 
 We analyze the spatial
 dependence of the heavy quark potential later and compare it against
 the other known forms of the potentials in the forthcoming sections. 
To that end, we employ the Debye mass computed from the effective
 fugacity quasi-particle model (EQPM)~\cite{chandra1,chandra2} and compare
 all the predictions with Debye mass obtained in HTL and Lattice QCD
 computations. Let us now proceed to discuss the EQPM and Debye mass below.
 
\subsection{The Debye mass from a quasi-particle picture of hot QCD} 
The Debye mass, $m_D$,  in QCD is generically non-perturbative and 
gauge invariant~\cite{arnold} unlike QED. The Debye mass in
leading-order in QCD coupling at high temperature has been 
 known from long time and is perturbative in nature
 \cite{shur}. In a work in the past,  Rebhan~\cite{rebh} defined $m_D$ by 
 seeing the relevant pole of the static quark propagator instead of
  the zero momentum limit of  the time-time component of the
 gluon self-energy.   The $m_D$ thus obtained is seen to be gauge independent.
 This follows from the fact that the pole of the self-energy is independent
 of choice of gauge. In their work, Braaten and Nieto~\cite{braaten}
 calculated the  $m_D$ for the QGP at high temperature to the
 next-to-leading-order (NLO) in QCD coupling from the correlator of two
 Polyakov loops (this agrees to the HTL result ~\cite{rebh}). Arnold
 and Yaffe~\cite{arnold} pointed out that the contribution of
 $O(g^2T)$ to the Debye mass in QCD  needs the knowledge of the
 non-perturbative physics of confinement of magnetic charge. They further
 argued that a perturbative definition of the Debye mass as a pole of
 gluon propagator no longer holds. Importantly, in lattice QCD, the definition of $m_D$ 
 itself, encounters difficulty due to the fact that unlike QED  the electric field correlators are not gauge invariant in QCD~\cite{ybu}. 
 To circumvent this problem, the approaches based on  effective theories obtained by dimensional reduction~\cite{ybu_3}, spatial correlation functions of gauge invariant meson correlators~\cite{ybu_4}, and 
the behavior of the color singlet free energies~\cite{ybu_57} have been proposed. In the concern, in a very recent attemp t by Burnier and Rothkopf~\cite{ybu} a 
gauge invariant mass has been defined from a complex static in medium  heavy-quark potential obtained from lattice QCD.
 
To capture all the interaction effects present in hot QCD equations
 of state in terms of non-interacting quasi-partons ( quasi-gluons and
 quasi-quarks), several attempts have been made. These  quasi-partons are
 nothing but the thermal excitations of the interacting quarks and gluons.
 We can categerize them as, (i) effective mass models\cite{vgolo,kampfer}, (ii) effective mass models with Polykov loop~\cite{polya}, (iii)
  models based on PNJL and NJL~\cite{pnjl}and (iv) effective fugacity model~\cite{chandra1,chandra2}.
 In QCD, the quasiparticle model is a phenomenological model which
 is widely used to describe the non-ideal bahavior of QGP near the phase
 transition point. The system of interacting massless quarks
 and gluons can be effectively described as an ideal gas of
 'massive' noninteracting quasiparticles in quasiparticle model.
 The mass of these quasiparticles is temperature-dependent and arises
 because of the interactions of quarks and gluons with the
 surrounding matter in the medium. These quasiparticles retain the
 quantum numbers of the real particles i.e., the quarks and
 gluons~\cite{sri}.
 \begin{figure*}
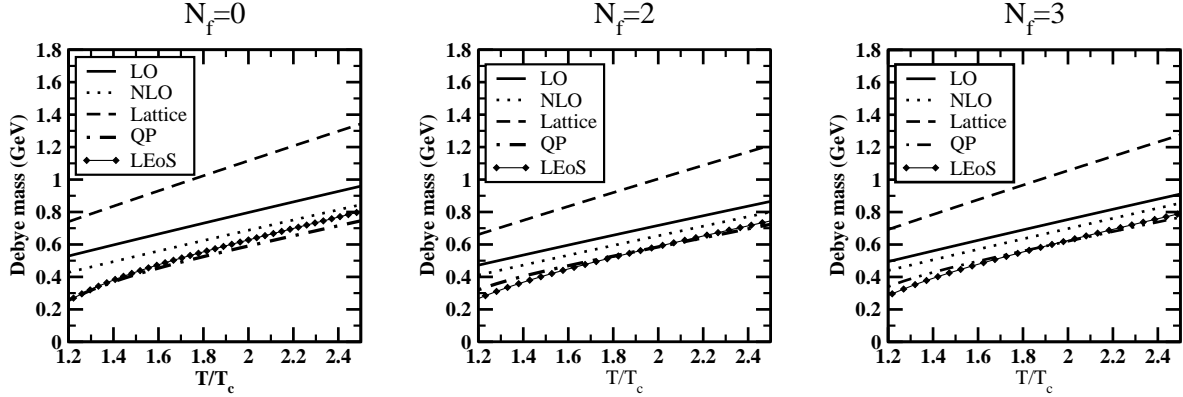

\vspace{3mm}
\includegraphics[scale=.45]{set1_d_QP_nf0.eps}
\hspace{5mm}
\includegraphics[scale=.45]{set1_d_QP_nf2.eps}
\hspace{5mm}
\includegraphics[scale=.45]{set1_d_QP_nf3.eps}
\caption{Debye mass verse temperature ($T/T_c$) for quasi particle (QP), lattice EoS, lattice parametrized, Next-to-leading order and leading order cases, when we used the 
fugacity equation of state EoS 1.Left panel represents the pure gluonic case, 
middle  and right panel represents 2-flavour and 3-flavour respectively.}
\end{figure*}

 Here, we consider
 the quasi-particle description \cite{chandra1,chandra2} of $O(g^5)$ hot
 QCD \cite{zhai,arnold} and $O(g^6\ln(1/g)$ hot QCD EoSs \cite{kajantie}, we call them 
 EoS1 and EoS2 respectively. We further consider the  lattice  QCD EoS ~\cite{lattice_baz} in terms of its quasi-particle description, we denote it as LEoS. Although there are more recent lattice 
 results with improved lattice actions and more refined lattice~\cite{lattice_hotqcd,lattice_fodor}, but to update the current model requires pure glue results for the trace anomaly with the same lattice set-up.
  Therefore,  such attempts are beyond the scope of the present
 work. We intend to  explore these possibilities in near future.
 
The equilibrium distribution function is written in the form given below:
\begin {equation}
\label{eq_1}
f_{g,q}=\frac{z_{g,q}\exp(-\beta p)}{\bigg(1\mp z_{g,q}\exp(-\beta p)\bigg)}.
\end {equation}
where $g$ stands for quasi-gluons, and $q$ stands for quasi-quarks. 
$z_g$ is the quasi-gluon effective fugacity and $z_q$ is quasi-quark
 effective fugacity. These distribution functions are isotropic
 in nature. These fugacities should not be confused with any
 conservation law (number conservation) and have merely been
 introduced to encode all the interaction effects at high temperature
 QCD. Both $z_g$ and $z_q$ have a very complicated temperature
 dependence and asymptotically reach to the ideal value unity
\cite{chandra2}. The  temperature dependence $z_g$ and $z_q$  fits well to the 
form given below,
\begin {equation}
\label{eq2}
z_{g,q}=a_{q,g}\exp\bigg(-\frac{b_{g,q}}{x^2}-\frac{c_{g,q}}{x^4} -\frac{d_{g,q}}{x^6}\bigg).
\end {equation}
(Here $x=T/T_c$ and $a$, $b$ and $c$ and $d$ are fitting parameters), for both EoS1 and EoS2.

The Debye mass, $m_D$ is defined in
 terms of the equilibrium (isotropic) distribution function as, 
\begin {equation}
\label{debye}
 m_D^2 \equiv -g^2 
 \int  \frac{{\rm d}^3\bar{\vec{p}}}{(2\pi)^3} \, 
 \frac{{\rm d}f_{eq}(\bar{p})}{{\rm d}\bar{p}}.
\end {equation}
where, $f_{eq}$ is taken to be a combination of ideal Bose-Einstein
 and Fermi-Dirac distribution functions as~\cite{rebhan}, and is given by:
\begin {equation}
\label{eq8}
f_{eq}=2 N_c f_{g}(\vec{p}) + 2 N_f (f_q(\vec{p}) + f_{\bar{q}}(\vec{p})).
\end {equation}
 Since, we are dealing with the QGP system with vanishing baryon density, therefore, 
$f_{q}=f_{\bar{q}}$ (here, $f_g$ and $f_q$ are the quasi-parton thermal distributions given in Eq. ({\ref{eq_1}})).
This combination of $f_{eq}$  leads to the leading order HTL expression 
($m_D^2=g^2(T) T^2 (N_c/3 +N_f/6)$) for the Debye mass in hot QCD. Here,
 $N_c$ denotes the number of colors and $N_f$ the number of flavors. 
 
Now, considering quasi-parton distributions,  we obtain, $m_D$ in the pure
 gluonic case:  
\begin{equation}
\label{notation}
m^2_D=g^2(T) T^2 \bigg(\frac{N_c}{3}\times\frac{6 PolyLog[2,z_g]}{\pi^2}\bigg)
\end{equation}
and full QCD:
\begin{eqnarray}
m^2_D &=& g^2(T) T^2 \bigg[
\bigg(\frac{N_c}{3}\times\frac{6 PolyLog[2,z_g]}{\pi^2}\bigg)\nonumber\\&&
+{\bigg(\frac{N_f}{6}\times\frac{-12 PolyLog[2,-z_q]}
{\pi^2}\bigg)}\bigg].
\end{eqnarray}
Here, $g(T)$ is the QCD running coupling constant, $N_c=3$ ($SU(3)$) and $N_f$
 is the number of flavor, the function $PolyLog[2,z]$ having form,
 $PolyLog[2,z]=\sum_{k=1}^{\infty} \frac{z^k}{k^2}$. We get same expressions from the chromo-electric response
 functions in~\cite{chandra3} for the interacting QGP.
 
 \begin{figure*}
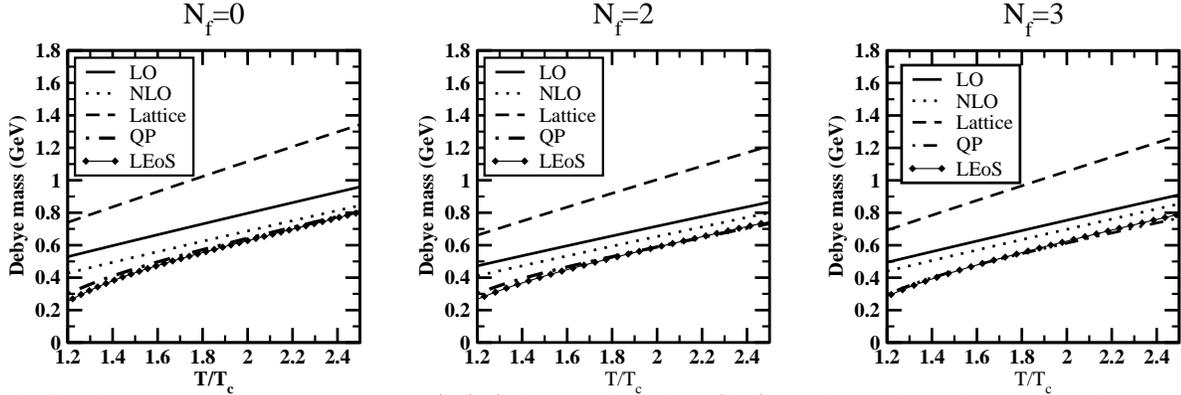

\includegraphics[scale=.45]{set2_d_QP_nf0.eps}
\hspace{5mm}
\includegraphics[scale=.45]{set2_d_QP_nf2.eps}
\hspace{5mm}
\includegraphics[scale=.45]{set2_d_QP_nf3.eps}
\vspace{-5mm}
\caption{Debye mass verse temperature ($T/T_c$) for quasi particle (QP), lattice EoS, lattice parametrized , and leading order (LO)  cases, when we used the
fugacity equation of state EoS 2.Left panel represents the pure gluonic case,
middle  and right panel represents 2-flavour and 3-flavour respectively.}
\end{figure*}

 The medium modified $m_D$ in terms of effective fugacities can be understood by relating it with the charge renormalization in 
 the medium. This could be done by defining the effective charges for the quasi-gluons and quarks as $Q_g$ and $Q_q$. 
 These effective charges are given by the equations:
\begin{eqnarray}
 Q^2_g&=&g^2 (T) \frac{6 PolyLog[2,z_g]}{\pi^2}\nonumber\\
 Q^2_q&=&g^2 (T)  \frac{-12 PolyLog[2,-z_q]}{\pi^2}.
\end{eqnarray}
Now the expressions for the Debye mass can be rewritten  as,
\begin{equation}
m^2_D=
\left\{\begin{array}{rcl}
Q^2_g T^2\frac{N_c}{3} &\mbox{for pure gauge,} &\\
T^2( \frac{N_c}{3} Q^2_g)+(\frac{N_f}{6} Q^2_q) &\mbox{for full QCD}&
\end{array} \right.
\end{equation}
Here, $\lbrace Q^2_g, Q^2_q\rbrace \le g^2(T)$ since it acquire the
 ideal value $g^2 (T)$ asymptotically. 
As mentioned earlier, the effective fugacities,  $ z_g $ and $ z_q$ are obtained for EoS1, EoS 2 and LEoS. The Debye mass with LEoS using our quasi-particle model model
 is seen closer to  that for EoS1 and EoS as compared to other cases. It is farthest as compared lattice Debye mass as the factor of $1.4$ in the definition of  of the lattice  Debye mass can not 
 be reproduced by perturbative/improved  perturbative QCD  or transport theory.

 The temperature dependence of the quasi-particle Debye mass, $m_D^{QP}$ in pure and full QCD with $N_f=2, 3$ is depicted in Fig. 1 and Fig. 2,  comparing it  with 
  the  LO and NLO in  HTL,  and lattice parameterized Debye masses which are denoted as $m_D^{LO}$ ,   and $m_D^{L}$ respectively. These various Debye masses have 
 the following mathematical expressions, 
  \begin{eqnarray}
  \label{dmm}
  m_D^{LO}&=& g(T) T \sqrt{\frac{N_c}{3}+\frac{N_f}{6}}, \nonumber\\
  m_D^{NLO}&=&  m_D^{LO}+\frac{N_c g^2(T) T}{4 \pi} \ln(\frac{m_D^{LO}}{g^2(T) T}),\nonumber\\
  m_D^{L}&=& 1.4 g(T) T,\nonumber\\
  m_D^{QP} &=&  g(T) T \bigg[\frac{2 N_c}{3\pi^2} PolyLog[2,z_g] \nonumber\\&&-\frac{2N_f}{\pi^2} PolyLog[2, -z_q] \bigg]^{\frac{1}{2}}.
  \end{eqnarray}
  For $g(T)$, we employ two expression for the running coupling in finite temperature QCD~\cite{laine_coupling}.
   Clearly, $m_D^{QP}$ is lowest among all other cases for the whole range of temperature considered here. The $m_D^{LO}$ is higher and $m_D^{NLO}$, and $m_D^{L}$ is largest among them for the 
  whole range of temperature.   From its temperature 
  dependence in Eq. (\ref{dmm}), it is straightforward to see that it will approach to the $m_D^{LO}$ asymptotically ($z_{g,q}\rightarrow 1$). These observations are holding true for all ($N_f=0, 2, 3 $) cases and for the EoS1 and ESO2.
 
\begin{figure*}
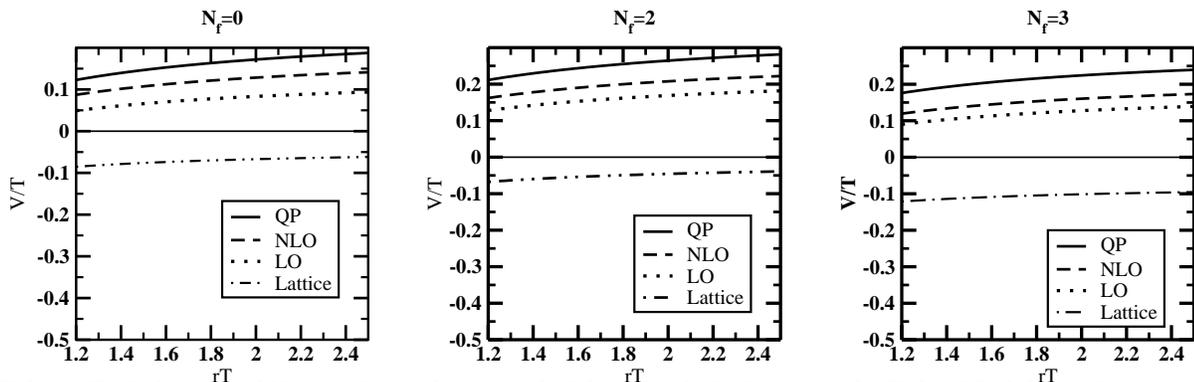

\includegraphics[scale=.45]{v_nf0_332_g5.eps}
\hspace{5mm}
\includegraphics[scale=.45]{v_nf2_332_g5.eps}
\hspace{5mm}
\includegraphics[scale=.45]{v_nf3_332_g5.eps}
\vspace{-5mm}
\caption{The behaviour of $V(r,T)/T$ as a function
of $r T$ for a fixed ($T/T_c=3.32$) for EoS1. The left panel represents the pure gluonic, 
middle and right panel represents 2-flavour and 3-flavour respectively.}
\vspace{25mm}
\end{figure*}

 \subsection{Heavy quark potential and quankonia Binding energies with EQPM}  

\subsubsection{The Heavy-quark Potential}
The heavy-quark potential given in Eq. ({\ref{eq6}}) is shown as a function of $r T$ for fixed $T/T_c$  for pure gluonic, 
$N_f=2$ and $N_f=3$ cases in Fig. 3 (for EoS1) and Fig. 4 (EoS2)  The expressions for the $m_D$ has been taken from Eq. (\ref{dmm}) and employed in 
the expression for the potential in Eq. ({\ref{eq6}}).  As expected the potential as a function of $r T$ is lowest with the $m_D^{L}$ and highest for the $m_D^{QP}$ for the fixed $T$ for the entire range of $r T$ (this just follows 
from the temperature dependence of the Debye mass {\it i. e.}, higher the Debye mass higher the screening). The similar observations are seen for $N_f=2$ and $3$ and for both EoS1 and EoS2.

\begin{figure*}
\vspace{40mm}
\includegraphics[scale=.45]{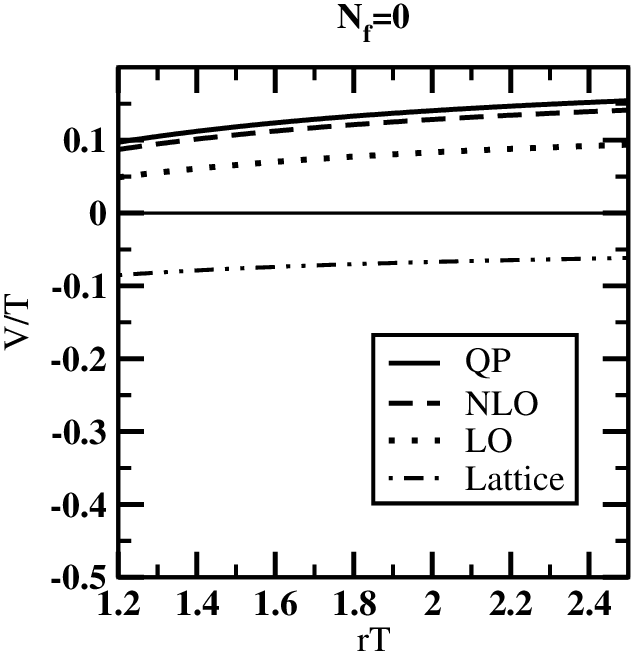}
\hspace{5mm}
\includegraphics[scale=.45]{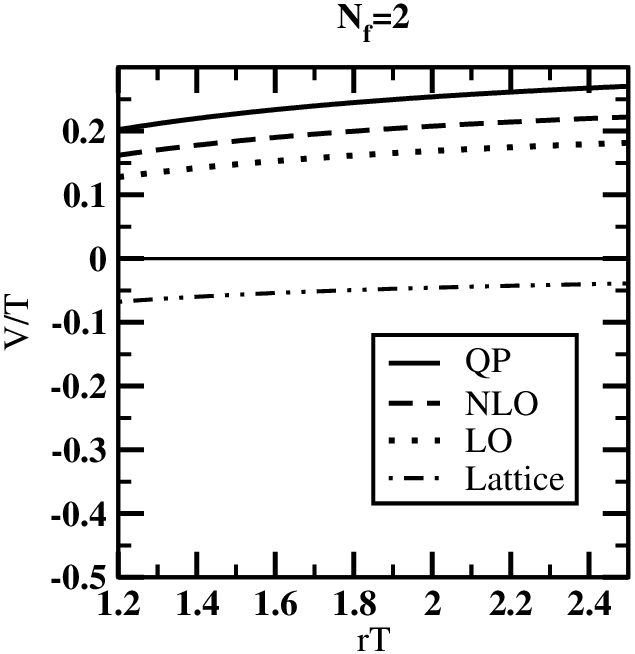}
\hspace{5mm}
\includegraphics[scale=.45]{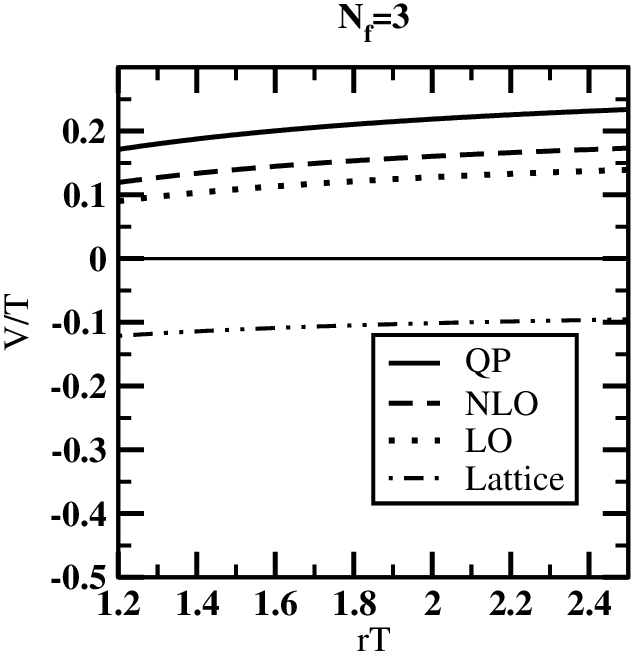}
\vspace{-5mm}
\caption{The behaviour of $V(r,T)/T$ as a function
of $r T$ for a fixed ($T/T_c=3.32$) for EoS2. The  left panel represents the pure gluonic, 
middle and right panel represents 2-flavour and 3-flavour respectively.}
\vspace{20mm}
\end{figure*}

\begin{figure*}
\vspace{10mm}
\includegraphics[scale=.55]{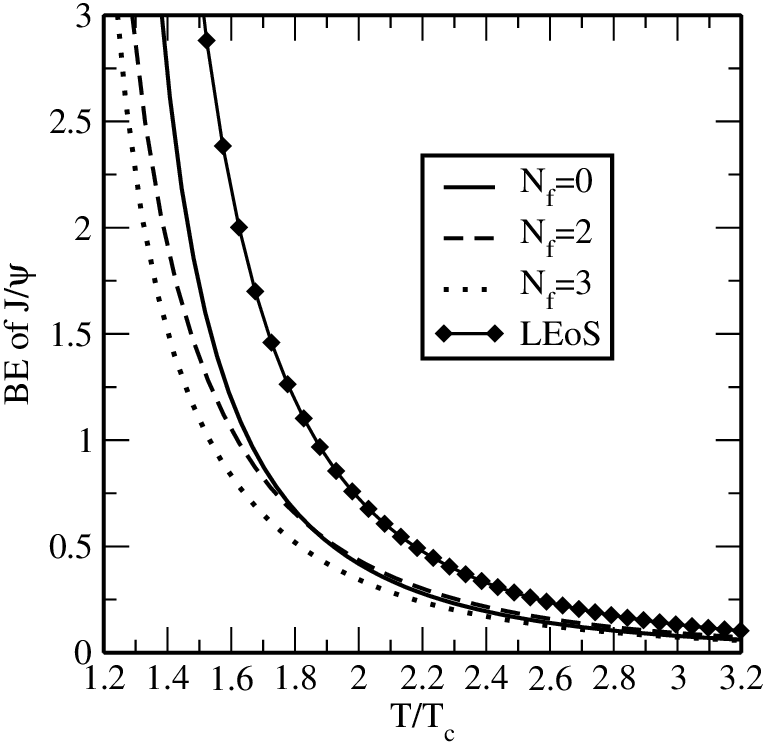}
\hspace{5mm}
\includegraphics[scale=.55]{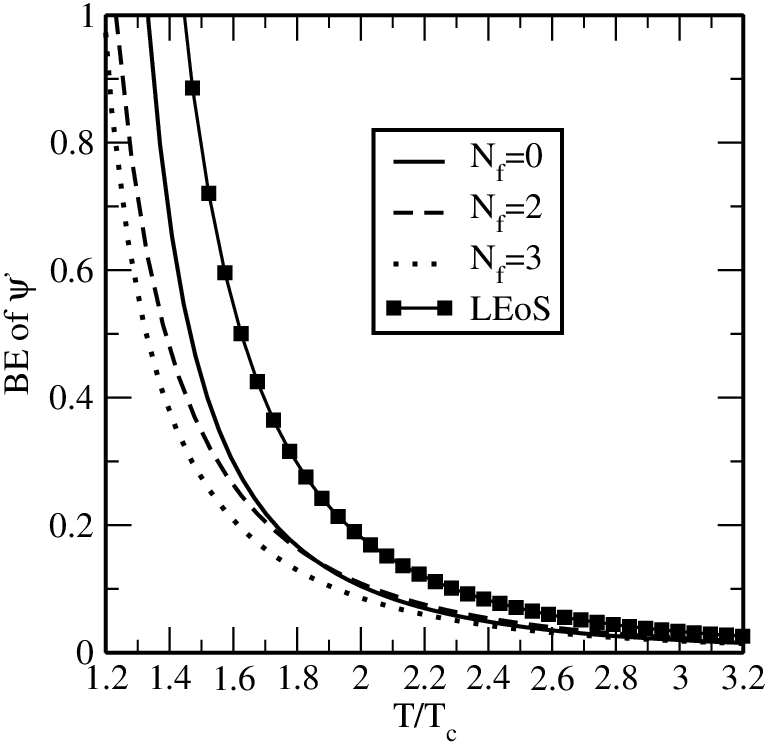}
\vspace{-5mm}
\caption{Dependence of binding energy(in $GeV$) of (a) $J/\psi$ and (b) 
$\psi^\prime$ on temperature $T/T_c$ with fugacity equation of state EoS1}.
\vspace{25mm}
\end{figure*}

\begin{figure*}
\vspace{10mm}
\includegraphics[scale=.55]{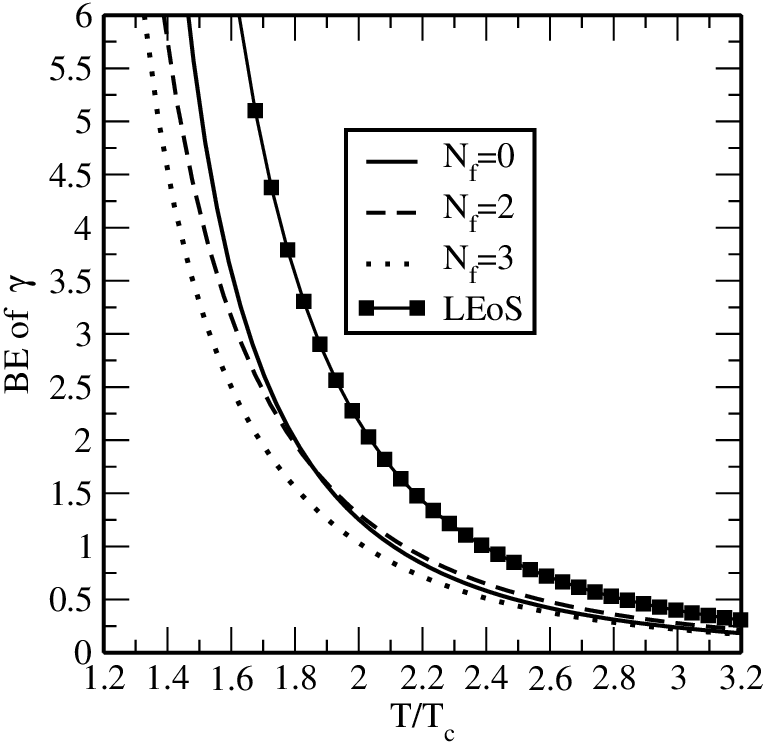}
\hspace{5mm}
\includegraphics[scale=.55]{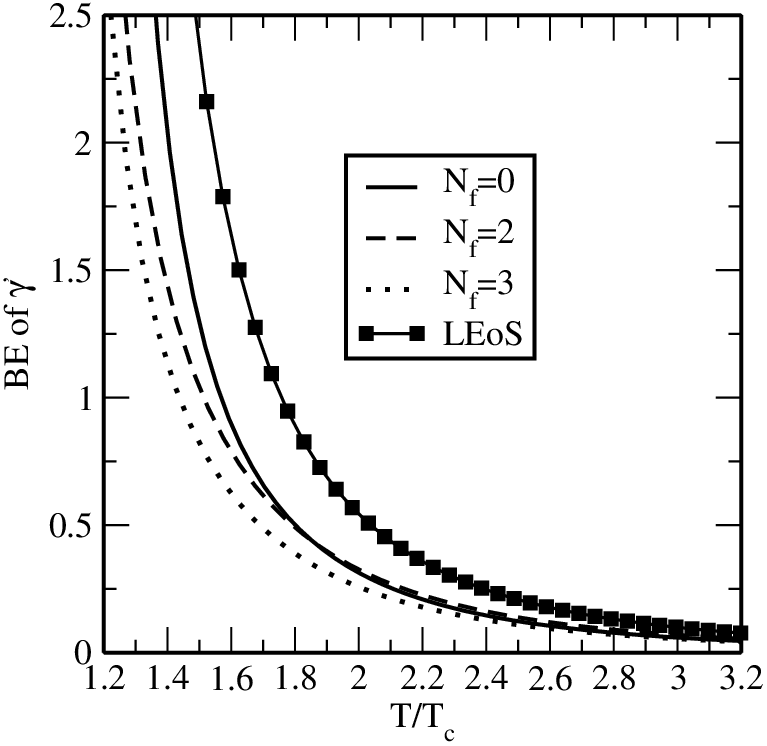}
\vspace{-5mm}
\caption{Dependence of binding energy(in $GeV$) of (a) $\Upsilon$ and (b) 
$\Upsilon^\prime$ on temperature $T/T_c$ with fugacity equation of state EoS1}.
\vspace{20mm}
\end{figure*}  
\subsubsection{The Binding Energies of $J/\Psi$ and $\Upsilon$ }  
To obtain the binding energy (BE) with heavy quark potential in Eq. ({\ref{eq6}}), we need to solve the Schr\"{o}dinger equation numerically with the full medium dependent  complex potential~\cite{kaka}.
Clearly, the binding energy will have both real and the imaginary parts.  One can take the intersection point of real and imaginary parts of the binding energies while plotting their temperature dependences to define the 
dissociation temperature of quarkonia state under consideration.  Another approach to look at the quarkonia dissociation is to first compute the thermal width of the given quarkonia from the imaginary part of the potential and equate it with the 
twice of the binding energies (real part). We follow the latter approach to estimate the  dissociation temperatures. Therefore, we shall mostly concentrate on the the real part of the binding energies and thermal width of the quarkonia.

In the limiting case 
discussed earlier, the real part of the medium modified  potential resembles to the hydrogen atom problem~\cite{matsui}.
 The solution of the Schr\"{o}dinger equation gives the eigenvalues for the 
ground states and the first excited states in charmonium
 ($\jpsi$, $\psi^\prime$ etc.) and bottomonium
 ($\Upsilon$, $\Upsilon^\prime$ {\it etc}.) spectra :
\begin{eqnarray}
\label{bind1}
E_n=-\frac{1}{n^2} \frac{m_Q\sigma^2}{m^4_D},
\end{eqnarray}
where $m_Q$ is the mass of the heavy quark. 

In the present case, we solve the Schr\"{o}dinger equation with full potential and obtain the binding energies. The temperature dependence of the binding energies 
are shown in  Figs. 5-8. For our analysis here, we consider $J/\Psi$, $\Psi^{\prime}$ binding energies  with EoS 1 and EoS 2 as a function of temperature  in  Fig. 5 and Fig. 7 respectively.
  On the other hand for  $\Upsilon$ and $\Upsilon^\prime$  with these equations of state in Fig. 6 and Fig. 8.
  
 We have also plotted the LEoS estimates for BEs  of various quarkonia states based on its quasi-particle understanding along with prediction for EoS1 and EoS2.
  The BE in this case is largest as compared to $N_f = 0,2,3$ using EoS1 and EoS2 for the considered range of temperature.This observations is seen to be valid for not only 
  $J/\psi$, $\Psi'$ but also$\Upsilon$ and $\Upsilon'$ states.
In each of the cases, the behavior is shown for $N_f=0, 2$ and $3$. There are some interesting observations that could be made while having a closer look 
at the temperature dependence of the binding energies in each case. Comparing the $J/\Psi$ and $\Psi^{\prime}$ cases, we see that the binding energy is 
approaching to zero sharply in the later case. This roughly implies that the latter state will dissolve before the former one. The same statement could 
me made for $\Upsilon$ and $\Upsilon^\prime$  states {\it i. e.},  the former will dissociate later in temperature as compared to the latter state.
We shall see that these observations are indeed true while we estimate the dissociation temperature for these states later. Interesting, for the 
three cases ($N_f=0, 2, 3$) with either EoS 1 and EoS 2 , these predictions for the dissociation temperatures come out true.

\begin{figure*}
\vspace{10mm}
\includegraphics[scale=.55]{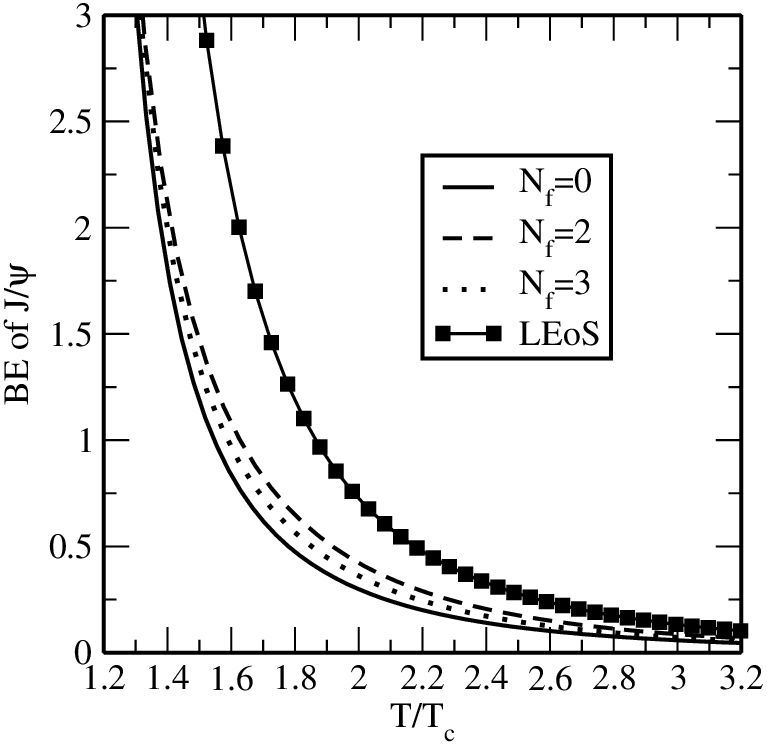}
\hspace{5mm}
\includegraphics[scale=.55]{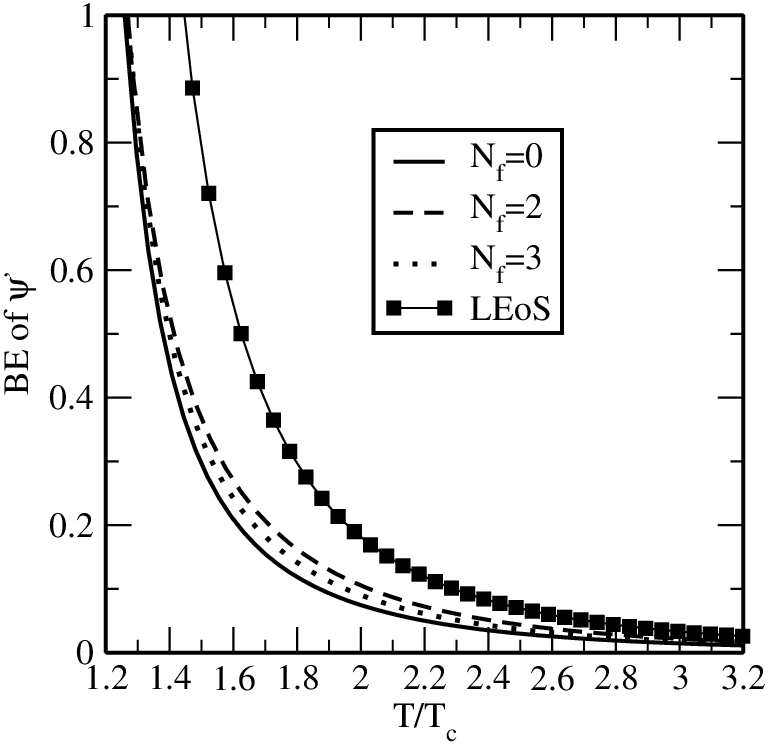}
\vspace{-5mm}
\caption{Dependence of binding energy(in $GeV$) of (a) $J/\psi$ and (b)
$\psi^\prime$ on temperature $T/T_c$ with fugacity equation of state EoS2}.
\vspace{25mm}
\end{figure*}

\begin{figure*}
\vspace{10mm}
\includegraphics[scale=.55]{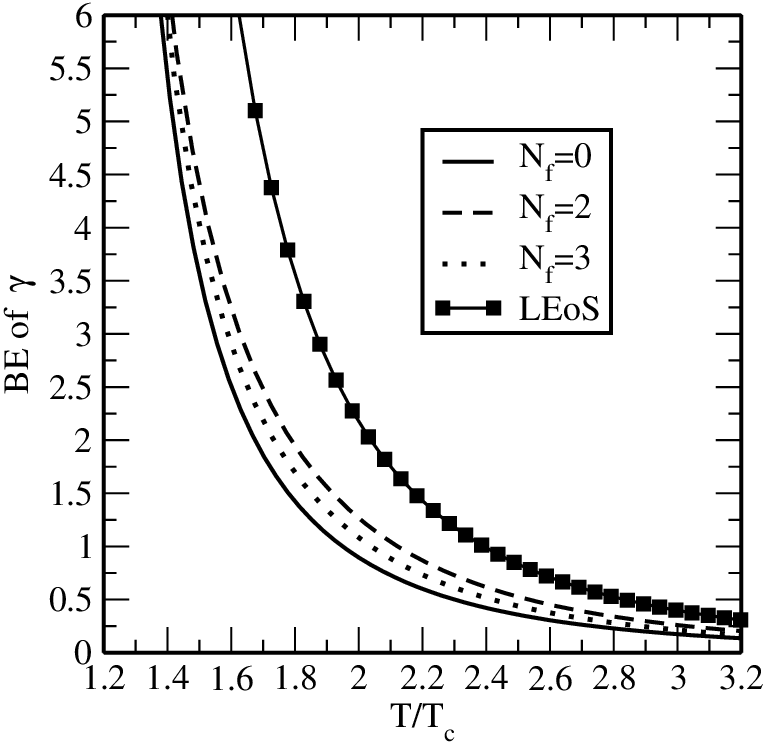}
\hspace{5mm}
\includegraphics[scale=.55]{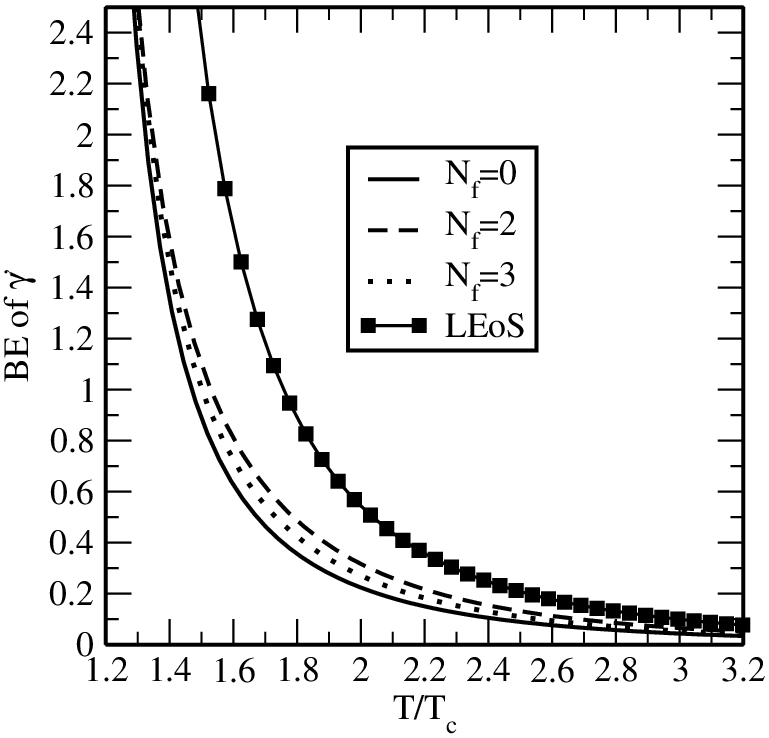}
\vspace{-5mm}
\caption{Dependence of binding energy(in $GeV$) of (a) $\Upsilon$ and (b)
$\Upsilon^\prime$ on temperature $T/T_c$ with fugacity equation of state EoS2}.
\end{figure*}

Let us now proceed to the computation of the dissociation temperatures for the above mentioned quarkonia bound states. To that end, we need  to compute the 
imaginary part of the heavy-quark potential and thus estimate the thermal width.

\section{The complex inter-quark potential}
Here, we discuss how to obtain the the complex inter-quark potential. The real part of the potential will be same as 
Eq.  (\ref{eq6}). We follow the similar procedure to obtain the imaginary part of the potential as discussed below.
To obtain the imaginary part of the inter-quark potential, we first need to 
obtain the imaginary part of the  symmetric self energy in the static limit. This can be done by  obtaining 
the imaginary part of the HTL propagator which represents the inelastic scattering
of an off-shell gluon to a thermal gluon~\cite{nora1,const2,Laine:2007qy,Escobedo:2008sy}.
The imaginary part of the potential plays crucial role in weakening 
the bound state peak or transforming it to mere threshold enhancement and eventually in 
dissociating it  (finite width ($\Gamma $) for the resonance 
peak in the spectral function, is estimated from the imaginary part of the potential which, in turn, determines the 
dissociation temperatures for the respective quarkonia). This sets the dissociation criterion, {\it i. e.}, it is expected 
 to occur while the (twice) binding energy  becomes equals the width
 $\sim \Gamma$~\cite{mocsy,Burnier:2007qm}. The equality will  do the quantitative determination of the dissociation temperature.
 
To obtain the imaginary part of the potential in the QGP medium,  
the temporal component of the symmetric propagator in the static limit has been considered as~\cite{laine},
\begin{equation}
Im D^{00}_{F(iso)}(0,k)=\frac{-2\pi T m_D^2}{k(k^2+m_D^2)^2}.
\label{isof}
\end{equation}
The same  expression Eq. (\ref{isof}) could also be obtained for 
partons with space-like momenta ($\omega^2 < k^2$) from the retarded 
(advanced) self energy~\cite{Dumitru:2009fy}  using the 
relation~\cite{const2,laine}: 
\begin{eqnarray}
\ln\frac{\omega+k\pm i\epsilon}{\omega-k\pm i\epsilon}=
\ln | \frac{\omega+k}{\omega-k}| \mp i \pi 
\theta(k^2-{\omega}^2) ~.
\end{eqnarray}
The imaginary part of the symmetric propagator Eq. (\ref{isof}) leasds to the 
the imaginary part of the dielectric function in the QGP medium as:
\begin{eqnarray}
\frac{1}{\epsilon (k)}= -\pi T m_D^2 \frac{k^2}{k(k^2+m_D^2)^2}.
\end{eqnarray}

Afterwards, the imaginary part of the in medium potential is easy to obtain
 owing the definition of the potential~Eq. (\ref{eq3}) as mentioned
 in~\cite{utt}:
\begin{eqnarray}
Im V(r,T)&=&-\int \frac{d^3\mathbf{k}}{(2\pi)^{3/2}}
(e^{i\mathbf{k} \cdot \mathbf{r}}-1)\nonumber\\
&&\times \left(-\sqrt{\frac{2}{\pi}}\frac{\alpha}{k^2}-\frac{4 \sigma}{\sqrt{2 \pi k^4}}\right) \frac{-\pi T m_D^2\ k}{(k^2+m_D^2)^2}\nonumber\\&&
\equiv Im V_{1}(r,T)+Im V_{2}(r,T)~,
\end{eqnarray}
where $Im V_{1} (r,T)$ and $Im V_{2}
(r,T)$ are the imaginary parts of the potential due to the 
medium modification to the short-distance and 
long-distance terms, respectively:
\begin{eqnarray}
Im V_{1}(r,T)&=&-\frac{\alpha}{2\pi^{2}}\int d^{3}\mathbf{k} 
(e^{i\mathbf{k} \cdot \mathbf{r}}-1)\nonumber\\&&\times
\left[\frac{\pi T m_D^2}{k(k^2+m_D^2)^2}\right],\\
Im V_{2}(r,T)&=&-\frac{4\sigma}{({2\pi})^2}\int \frac{d^3 {\bf k}}
{(2\pi)^{3/2}}(e^{i {\bf k} \cdot {\bf r} }-1)\nonumber\\&&\times
\frac{1}{k^2}
\left[\frac{\pi T m_D^2}{k(k^2+m_D^2)^2}
\right].
\end{eqnarray}
After performing the integration, the contribution due to the short-distance
term to imaginary part becomes (with $z=k/m_D$)
\begin{eqnarray}
Im V_{1}({\bf r},T)&=&2\alpha T\int_0^{\infty} \frac{zdz}{(z^2+1)^2}
\left(1-\frac {\sin{z\hat r}}{z\hat r}\right)\nonumber\\
&\equiv &\alpha T\phi_0(\hat{r}),
\label{imiso1}
\end{eqnarray}
and the contribution with the non-zero  string tension becomes:

\begin{eqnarray}
Im V_{2}(r,T) &=&\frac{4\sigma T}{m_D^2}\int_0^{\infty} 
\frac{dz}{z(z^2+1)^2}
\left(1-\frac {\sin{z\hat r}}{z\hat r}\right)\nonumber\\
&\equiv &\frac{2\sigma T}{m_D^2}\psi_0(\hat{r})~,
\label{imiso2}
\end{eqnarray}
where the functions, $\phi_0(\hat{r})$ and $\psi_0(\hat{r})$ 
at leading-order in $\hat{r}$ are
\begin{eqnarray}
\label{eq24}
\phi_0(\hat{r})&=& \left(-\frac{{\hat{r}}^2}{9}
\left(-4+3\gamma_{E}+3\log\hat{r}\right)\right).
\end{eqnarray}

\begin{eqnarray}
\label{eq25}
\psi_0(\hat{r})&=&\frac{\hat r^2}{6}+\left(\frac{-107+60\gamma_E 
+60\log(\hat r)}{3600}\right)\hat r^4+O(\hat r^5).\nonumber\\
\end{eqnarray}
In the short-distance limit ($ \hat{r} \ll 1$), both the 
contributions, at the leading logarithmic order, reduce to
\begin{eqnarray}
&&Im V_{1}(r,T)=\alpha T\frac{{\hat r^2}}{3}\log(\frac{1}{\hat r}),\\
\label{im1}
&&Im V_{2}(r,T)=-\frac{2\sigma T}{m_D^2}\frac{{\hat r^4}}{60}
\log(\frac{1}{\hat r}).
\label{im2}
\end{eqnarray}
Therefore, the sum of Coulomb and string tension dependent terms
leads to the  the imaginary part 
of the potential:
\begin{eqnarray}
Im V (r,T)=T\left(\frac{\alpha {\hat r^2}}{3}
-\frac{\sigma {\hat r}^4}{30m_D^2}\right)\log(\frac{1}{\hat r}).`
\label{imis}
\end{eqnarray}
One thus immediately observes that for small distances the imaginary part 
vanishes and its magnitude is smaller as compared to the  case with  only 
the Coulombic term~\cite{Dumitru:2009fy}. The effect of non-perturbative contribution coming from the
string terms, thus, reduces the  width of the 
resonances in thermal medium.
The imaginary part of the potential above, provides an estimate 
for the width ($\Gamma$) for a resonance state. The width $\Gamma$ can be computed in  first-order perturbation,
while folding the imaginary part of the potential with the unperturbed 
(1S) Coulomb wave function as:
\begin{eqnarray}
\Gamma =\left(1+\frac{3\sigma }{\alpha m_Q^2}\right)~\frac{4T}{\alpha }{\frac{m_D^2 }{m_Q^2}}~\log\frac{\alpha m_Q}{2m_D}.
\end{eqnarray}

\begin{figure*}
\includegraphics[scale=.80]{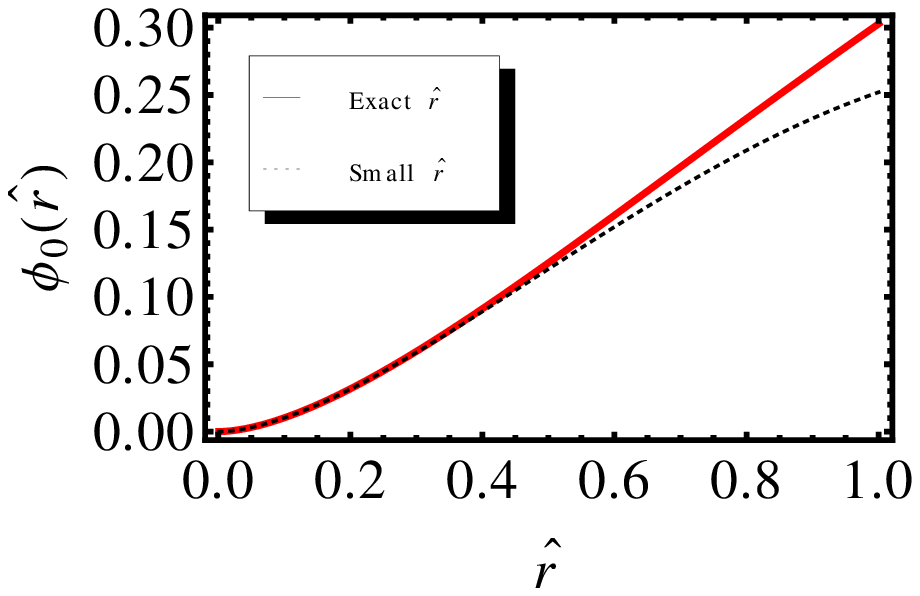}
\includegraphics[scale=.72]{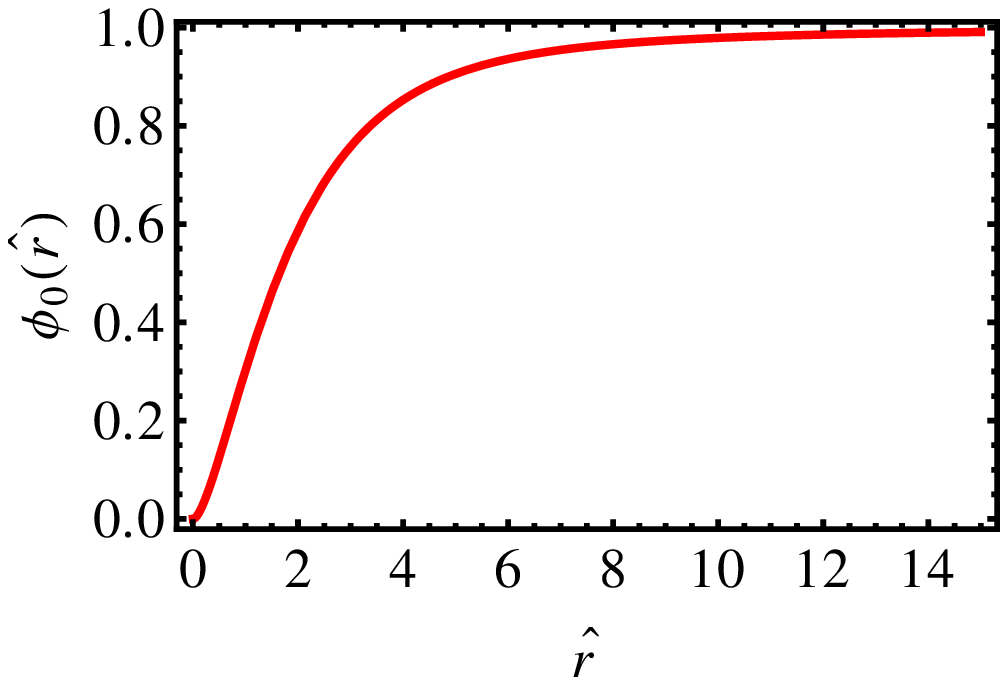}
\vspace{-5mm}
\caption{Dependence of  $\phi_0(\hat{r})$ on $\hat{r}$. The lower panel shows the 
comparison between the small $\hat{r}$ approximation and exact one.  }
\label{ps11}
\vspace{25mm}
\end{figure*}

\begin{figure*}
\includegraphics[scale=.75]{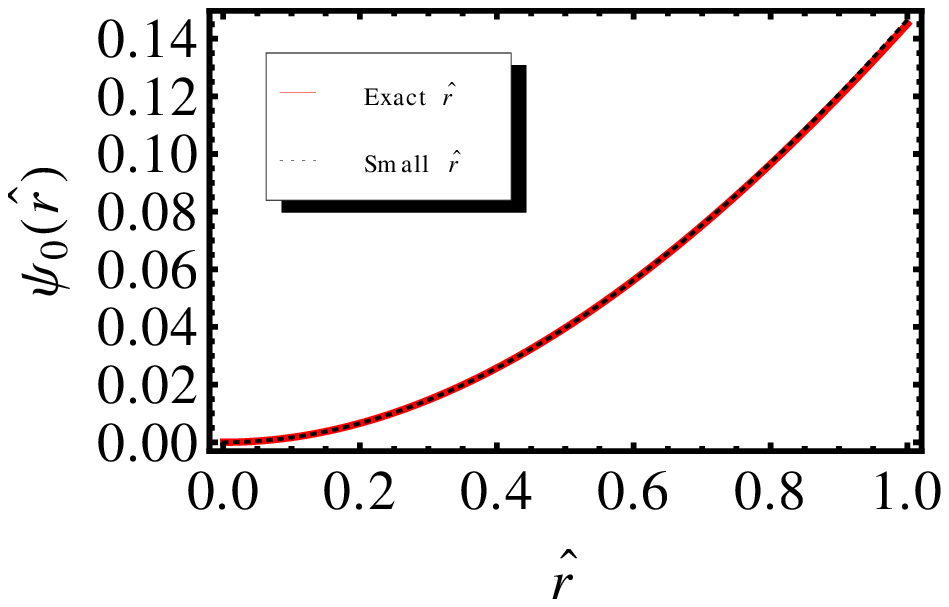}
\includegraphics[scale=.85]{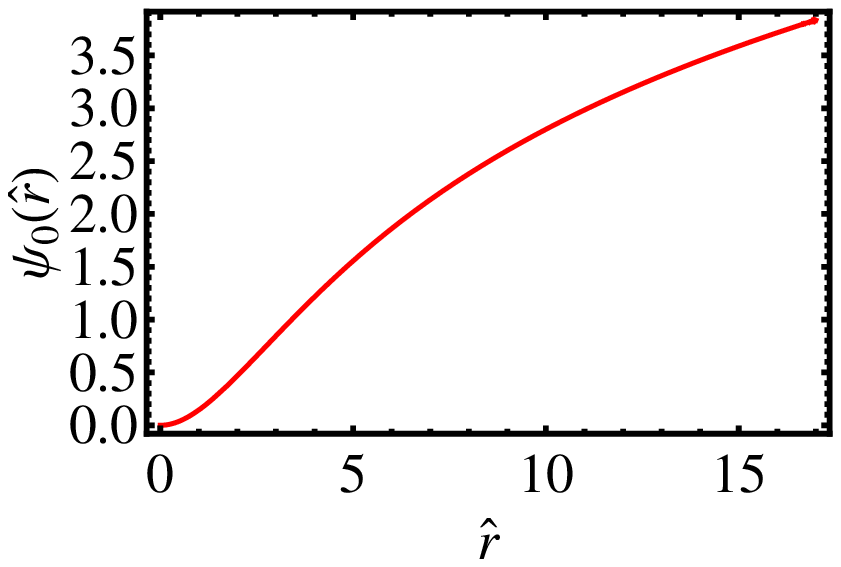}
\vspace{-5mm}
\caption{ Dependence of  $\psi_0(\hat{r})$ on $\hat{r}$. The figure in the left shows the 
comparison between the small $\hat{r}$  approximation and exact results. The right figure depicts the exact results until $\hat{r}=17$ beyond which fluctuations start showing up and grow rapidly.  }
\label{ps111}
\end{figure*}

It is possible to solve the integral for the functions  $\phi_0 (\hat{r})$ and $\psi_0 (\hat{r})$
in the right hand side of the Eq. (\ref{imiso2}) exactly. The compact mathematical expressions are 
presented in the Appendix. The behavior of these functions as a function of $\hat{r}$ is depicted in 
Figs. 9 and  10 where we have compared the small $\hat{r}$ behavior in Eq. (\ref{im2}) with approximate result
in Eqs. (\ref{eq24}) and (\ref{eq25}) along with results for larger $\hat{r}$.
 Clearly the approximation works fantastically well of 
$\hat{r}<1$ for $\phi_0 (\hat{r})$ and better for $\psi_0 (\hat{r})$. The behavior at large $\hat{r}$ is crucial to understand the fate of 
higher (excited) states of quarkonia. The analytic estimate for $\psi_0(\hat{r})$ based on the expression quoted in the appendix is well behaved until $\hat{r}\leq 16-17$.
For $\hat{r}> 17$ the functions $\psi_0(\hat{r})$ show large fluctuations that grow rapidly for larger $\hat{r}$. Therefore, in that region, we perhaps can not utilize it for phenomenological purposes.

\subsection{The dissociation temperatures for heavy quarkonia}
There are two criteria for the dissociation of quarkonia 
bound state in the QGP medium that are  under consideration here. 
The first one is the dissociation of a given quarkonia bound state by the 
thermal effects alone. On the other hand, the second criterion is based on the 
dissolution of a given quarkonia state while its thermal width is overcomed by the twice of the 
real part of the binding energy. We shall employ both of them one by one below and present the comparison of the 
quantitative estimates of the dissociation temperatures.

\subsubsection{Dissociation by thermal effects}
Dissociation of a quarkonia bound state in a thermal QGP medium will occur whenever 
the binding energy (BE), $E_B$ of the said state will fall below the 
mean thermal energy of a quasi-parton. In such situations the thermal effect can dissociate the 
quakonia bound state.

\begin{table}
\label{t1}
\caption {Lower(upper) bound on the dissociation temperature($T_D$) for the 
quarkonia states (in units of $T_c$)for using fugacity parameters of EoS 1}.
\centering
\begin{tabular}{|l|l|l|l|l|}
\hline
State &Pure QCD & $N_f=2$&$N_f=3$\\
\hline\hline
$\jpsi$&1.6(1.9) & 1.6(2.1) & 1.5(2.0) \\
\hline
$\psi'$&1.3(1.5) & 1.3(1.6) & 1.3(1.5)  \\
\hline
$\Upsilon$&1.9(2.4) & 2.1(2.6) & 2.0(2.5) \\
\hline
$\Upsilon'$&1.5(1.8) & 1.6(1.9) & 1.5(1.9) \\
\hline
\end{tabular}
\end{table} 

To obtain the lower bound of the dissociation temperatures of the various quarkonia states, the 
 (relativistic) thermal energy of the partons will be $3\  T$. On the other hand,  the upper bound of the
 dissociation temperature ($T_D$) is obtained by considering the mean thermal energy to be 
 $T$. The dissociation is supposed to occur whenever,
\begin{equation}
\label{tdiss}
E_B (T_D)= 3 T_D\  \textbf{or}\    T_D.
\end{equation}
 While solving for the $E_B$, the string tension ($\sigma$) is taken as $0.184$ ${\rm{GeV}^2}$,
and critical temperatures ($T_c$) are considered as   $270 MeV$, $203 MeV$
 and $197 MeV$ for pure, 2-flavor and 3-flavor QCD at high temperature 
 for  both the equations of state. The binding energies are shown as a function of temperature in 
 earlier plots. The dissociation temperatures  for the ground state  and the
 first excited state of $c\bar{c}$  ($J/\Psi$ and $\Psi^\prime$)  and $b\bar{b}$ sates ($\Upsilon$ and $\Upsilon^\prime$) are presented in 
 Table I and III while considering two different criteria of quarkonia dissociation.

\begin{table}
\label{t2}
\caption {Lower(upper) bound on the dissociation temperature($T_D$) for the
quarkonia states (in units of $T_c$)for using fugacity parameters of EoS 2}.
\centering
\begin{tabular}{|l|l|l|l|l|}
\hline
State &Pure QCD & $N_f=2$&$N_f=3$\\
\hline\hline
$\jpsi$& 1.5(1.8) & 1.7(2.0) & 1.6(1.9) \\
\hline
$\psi'$& 1.2(1.4) & 1.3(1.6) & 1.3(1.6) \\
\hline
$\Upsilon$& 1.8(2.2) & 2.0(2.6) & 2.0(2.5) \\
\hline
$\Upsilon'$& 1.4(1.7) & 1.6(1.9) & 1.6(1.9) \\
\hline
\end{tabular}
\end{table}

\subsubsection{Overcoming thermal width of the resonance by the binding energy}
Whenever the thermal width,  $\Gamma$ of the a given quarkonium is as large as  twice  the binding energy (real part)  the given quarkonia state will dissolve~\cite{utt}
\begin{table}
\label{t3}
\caption {The dissociation temperature($T_D$) for the
quarkonia states (in units of $T_c$)for using fugacity parameters of EoS 1, when thermal width =2\ BE }.
\centering
\begin{tabular}{|l|l|l|l|l|}
\hline
State &Pure QCD & $N_f=2$&$N_f=3$\\
\hline\hline
$\jpsi$& 1.8 & 2.0 & 1.9 \\
\hline
$\psi'$& 1.6 & 1.8 & 1.8 \\
\hline
$\Upsilon$& 2.6 & 2.8& 2.2 \\
\hline
$\Upsilon'$& 2.1& 2.2 & 2.1 \\
\hline
\end{tabular}
\end{table}

\begin{table}
\label{t4}
\caption {The dissociation temperature($T_D$) for the
quarkonia states (in units of $T_c$)for using fugacity parameters of EoS 2, when thermal width =2\ BE }.
\centering
\begin{tabular}{|l|l|l|l|l|}
\hline
State &Pure QCD & $N_f=2$&$N_f=3$\\
\hline\hline
$\jpsi$& 1.7 & 1.9 & 1.9 \\
\hline
$\psi'$& 1.5 & 1.7 & 1.7 \\
\hline
$\Upsilon$& 2.5 & 2.7& 2.6 \\
\hline
$\Upsilon'$& 2.0& 2.2 & 2.1 \\
\hline
\end{tabular}
\end{table}

 We applied the criteria for the $c\bar{c}$ bound states ($\jpsi$  and $\psi'$) and $b\bar{b}$ bound state ($\Upsilon$ and $\Upsilon'$). The quantitative
 estimates for the respective dissociation temperatures are enlisted in 
 Tables II and  IV.
 
 \begin{table}
\label{tabl5}
\caption {Lower(upper) bound on the dissociation temperature($T_D$) for the
quarkonia states for  2+1 flavour (in units of $T_c$) case while using the  fugacity parameters of the LEoS (second row). The third row records the estimates with  second criterion of the dissociation ( 2 \ BE $\equiv$ thermal width)
}
\centering
\begin{tabular}{|l|l|l|l|l|l|}
\hline
State &$\jpsi$ & $\psi'$&$\Upsilon$ &$\Upsilon'$\\
\hline\hline
LEoS & 1.9(2.3) & 1.5(1.8) & 2.3(2.8) & 1.8( 2.1) \\
\hline
LEoS &2.1 & 1.8 & 3.1 & 2.6 \\
\hline
\end{tabular}
\end{table}

Let us now analyze the quantitative estimates for $\jpsi$ and $\psi'$ dissociation temperatures for EoS1 equating the thermal width with the twice of the  BE.
The $\jpsi$ state is seen to dissociate at $T = 1.8T_{c}$ for $N_{f} = 0$ , $T = 2.0T_{c}$ for $N_{f} = 2$ and for $N_{f} = 3$ at $T = 1.9T_{c}$.
On the  $\Psi^\prime$ is seen to dissociate at $T = 1.6 T_{c}$ for $N_{f} = 0$ , $T = 1.8T_{c}$ for $N_{f} = 2$ 
and for $N_{f} = 3$ at $T = 1.8T_{c}$. On the other hand, for EoS2, $\jpsi$ is seen to dissociate at $T = 1.7T_{c}$ for $N_{f} = 0$ , $T = 1.9T_{c}$ for $N_{f} = 2$ and for $N_{f} = 3$ at $T = 1.9T_{c}$. $\Psi^\prime$ is seen to dissociate at $T = 1.5T_{c}$ for $N_{f} = 0$ , $T = 1.7T_{c}$ for $N_{f} = 2$ 
and for $N_{f} = 3$ at $T = 1.7T_{c}$ . As stated earlier (on the basis of temperature dependence of the BE ) $\Psi^\prime$ is seen to dissociate at lower temperatures as compared to $\jpsi$ for both the equations of state.

Similarly, for $\Upsilon$ and $\Upsilon^\prime$ dissociation temperatures are recorded in Table III and Table IV. The 
$\Upsilon$ state is seen to dissociate at $T = 2.6T_{c}$ for $N_{f} = 0$ , $T = 2.8T_{c}$ for $N_{f} = 2$ and for $N_{f} = 3$ at $T = 2.2T_{c}$ while employing EoS1 through the quasi-particle picture. 
On the other hand, $\Upsilon^\prime$ is seen to dissociate at $T = 2.1T_{c}$ for $N_{f} = 0$ , $T = 2.2T_{c}$ for $N_{f} = 2$ and for $N_{f} = 3$ at $T = 2.1T_{c}$ for the same EoS.
With  EoS2,  $\Upsilon$ is seen to dissociate at $T = 2.5T_{c}$ for $N_{f} = 0$ , $T = 2.7T_{c}$ for $N_{f} = 2$ and for $N_{f} = 3$ at $T = 2.6T_{c}$ and  $\Upsilon^\prime$ is seen to dissociate at $T = 2.0T_{c}$ for $N_{f} = 0$ , $T = 2.2T_{c}$ for $N_{f} = 2$ 
and for $N_{f} = 3$ at $T = 2.1 T_{c}$. Again,  we can see (on the basis of temperature dependence of the BE ) that $\Upsilon$ is seen to dissociate at higher temperatures as compared to $\Upsilon^\prime$ for both the equations of state.

The estimates for various quarkonia states under consideration with LEoS are quoted in 
Table V. The first row records the estimates for the case while  the  quarkonia dissociation has been  led by the average thermal energy of the $q/\bar{q}$.
On the other hand the second row captures estimates  while the BEs are  overcomed by the thermal width of quarkonia due to complex nature of the potential(inter-quark).
 The  upper bound obtained in row1 are closer to those with the latter criterion. On comparing the estimate only slightly different.
 
Comparing the numbers for the $T_D$  for various quarkonia states,  quoted in Table I and  Table III , we observe  that the quantitative estimates in Table III  are quite 
closer to the upper bound(NR) criteria. Note that the former estimates are based on the dissolution of a given quarkonia state by the mean thermal energy of the quasi-partons in the 
hot QCD/QGP medium, the latter one is  based on equating the  thermal width to the real part of the binding energy (twice).  Similar observations are obtained while comparing the estimates  from Table II  and Table IV.
Interesting, the numbers obtained by employing EoS1 and EoS2 with the latter criterion of quarkonia dissociation, the estimates are not very different from each other.

\section{Results and Discussion}
 The Hot QCD equations of state
 corresponding to interactions up to $O(g^5) $ and $O(g^6(ln1/g))$ in the
 improved perturbative QCD  can significantly impact the fate of quarkonia in 
 the QGP medium.  The medium modified
 form of the heavy quark-potential in which the medium modification
 causes the Debye screening of color charges, have been obtained by employing the 
 Debye mass obtained by utilizing the quasi-particle understanding of these 
 equations of state. This, in turn,  leads to the
 temperature dependent binding energies for the  $\jpsi$ and $\psi^\prime$. The binding energies 
 are seen to  decreases less sharply for pure gluonic case in comparison to full QCD medium. 
 Similar pattern have  been observed for the case of $\Upsilon$ and $\Upsilon^\prime$ states.
 
 To estimate the  dissociation temperature,  we consider two criteria {\it viz.} the dissociation by mean thermal energy 
 of the quasi-particles in the QGP medium and, the binding energy overcoming the thermal width of the 
 quakonia bound state. The upper and lower bound within the first criterion were obtained by 
 thermal energy  $T$ and $3T$, respectively. In numbers for the dissociation temperatures from both the criteria 
 are seen to be consistent with the recent predictions from the 
 recent quarkonium spectral function studies using a potential model. The effects of realistic EoS for the QGP have 
 significant impact on the binding energies and the dissociation temperatures for the various quarkonia states. 

\section{Conclusion and outlook}
In conclusion, we have studied the quarkonia dissociation in QGP in the
isotropic case  employing quasi-parton equilibrium distribution
 functions obtained from $O(g^5) $ and $O(g^6(ln1/g))$ hot QCD equations
 of state  and LEoS and medium modification to a heavy quark potential. We have found
 that medium modification causes a dynamical screening of color charge which,
in turn, leads to a temperature dependent of binding energy. We have
 systematically studied the temperature dependence of binding energy for the
 ground and first excited states of charmonium and bottomonium spectra in
 pure gluonic and full QCD medium. We have then determined the dissociation
 of heavy Quarkonium in hot QCD medium by employing the medium
 modification to a heavy quark potential and explore how the pattern
 changes for pure gluonic case and full QCD in the Debye mass.

 We intend to look for extensions of the present work in the case of 
 hydrodynamically expanding viscous QGP medium. Another, interesting direction would be 
 to couple the analysis to the physics of momentum anisotropy and instabilities in the early stages 
 of the heavy-ion collisions and its impact on the physics of heavy quarkonia dissociation and yields in 
 heavy-ion collisions.

\appendix
\section{Imaginary part of inter-quark potential}
It is possible to solve the integral in , Eq.(\ref{imiso1}) and Eq.(\ref{imiso2}) for real $\hat{r}$. The 
expression for the function $\phi^{0} (\hat{r})$  and $\psi^{0} (\hat{r})$ are obtained as:
\begin{eqnarray}
 \phi_{0}(\hat{r}) &=&1-\sqrt{\pi} G_{1,3}^{2,1} \left(\frac{\hat{r}^2}{4} \ {} \Bigg\vert \ {}{ {0}  \atop  {0 }, {1},
 {-\frac{1}{2}}}\right),
\nonumber\\
\psi_{0}(\hat{r})& =&\frac{1}{2|\hat{r}|}\Bigg(-6|\hat{r}| +4|\hat{r}|\gamma_{E} +4\hat{r}Log[|\hat{r}|]\nonumber\\ && 
+\Big(Ci\big(-i|\hat{r}|\big)+Ci\big(i|\hat{r}|\big)\Big)\nonumber\\ && \times
\bigg[|\hat{r}|\cosh\big(|\hat{r}|)-3\sinh\big(|\hat{r}|\big)\bigg]\nonumber\\ &&
+ 2Shi\big(|\hat{r}|\big)\bigg[3\cosh\big(|\hat{r}|)-\hat{r}|\sinh\big(|\hat{r}|\big)\bigg]\Bigg). \nonumber\\
\end{eqnarray}
Here, $ G$ is {\it MeijerG} function and
\begin{eqnarray}
Ci(z) &=& CosIntegral(z) = -\int_z^{\infty}\frac{\cos(t)}{t}dt,\nonumber\\
Shi(z) &=& SinhIntegral(z) = \int_0^{z}\frac{\sinh(t)}{t}dt\nonumber\\
\end{eqnarray}

\section*{Acknowledgement}
VKA acknowledge the UGC-BSR research start up grant No. {\bf F.30-14/2014 (BSR)}
New Delhi. VC would like to acknowledge DST, Govt. of India for the 
INSPIRE Faculty Award: {\bf IFA-13,PH-55}. We record our sincere gratitude to 
the people of India for their generous support for the research in basic sciences.

\end{document}